\documentclass[aps,preprint,showpacs]{revtex4}
\usepackage{graphicx}

\begin{document}
\title{Chaotic Emission from Electromagnetic Systems Considering
Self-Interaction} 
\author{Fernando Kokubun} 
\affiliation{Departamento de F\'\i sica, Universidade Federal 
do Rio Grande, 96201-900 Rio Grande, RS, Brazil} 
\author{ Vilson T. Zanchin} 
\affiliation{ Departamento de
F\'{\i}sica, Universidade Federal de Santa Maria, 97119-900 Santa
Maria, RS, Brazil}

\begin{abstract}
The emission of electromagnetic waves from a system described by the
H\'enon-Heiles potential is studied in this work.
The main aim being to analyze the behavior of the system when the damping term
is included explicitly into the equations of motion. Energy losses
at the chaotic regime and at the regular regime are compared.  The results
obtained here are similar to the case of gravitational waves emission,
as long we consider only the energy loss. The main difference being
that in the present work the energy emitted is explicitly calculated
solving the equation of motion  without further approximations.
It is expected that the present analysis may be useful when studying the
analogous problem of dissipation in gravitational systems.
\end{abstract}


\maketitle

\section{Introduction}

\subsection{Motivation}

The chief motivation of the present work is trying to better
understand the effects of damping forces (the radiation reaction
forces) in radiating systems undergoing chaotic motions. Our ultimate
interest is in gravitational systems, in particular, in the case of
radiating chaotic systems. However, due to the difficulties 
that usually arise during the numerical evolution of equations from
Einstein gravity, we start analyzing the electromagnetic 
analogous problem and shall use the experience acquired 
here to be able to circumvent those difficulties in future 
work dealing with gravitational systems.

In a classical field theory, the losses of energy and momenta due to
the presence of radiation reaction forces is of fundamental importance to
determine the physical properties of the system. Studies on this subject
have been done in electromagnetic systems since Maxwell has established the
foundations of the electromagnetic interaction, and in gravitational
systems just after Einstein has formulated the theory of general
relativity. Even though much progress have been done in both cases,
there are still some points to be clarified.

Recent works about emission of gravitational waves from chaotic
systems presented  several interesting features of such systems.
However, some important questions remained without answer.
Particularly, the influence of the damping term to the dynamics of a
chaotic system is not well understood.  A major difficulty in studying the
effects of radiation reaction  in the dynamics of a (chaotic) gravitational
system is the necessity of including higher order Post-Newtonian (PN) terms.
  Levin  \cite{levin2000} has shown that at $2.0$ PN
order, the two body problem with spin is chaotic, extending previous
study of Suzuki and Maeda \cite{suzuki}. Nevertheless, the effects of
a dissipation term become important only with inclusion of $2.5$ PN
order. So, in order to describe possible effects of chaotic emission
on the detection of gravitational waves, it becomes important to
consider higher order terms  (see for example the comments from
Cornish \cite{cornish} and Hughes \cite{hughes}).

It has been shown that the amount of energy carried away by
gravitational waves in a chaotic regime is smaller than 
in a regular regime \cite{kokubun} (see also
\cite{cornish,suzuki}). However, this result was obtained by brute
force method, because in Newtonian gravity, the emission of
gravitational waves is dynamically unimportant, and in these works the
energy emission was considered at Newtonian approximation. Thus,  knowing
the exact manner in which the emission of gravitational waves in a chaotic
system is affected by the damping term is still an open question. 
The way to find the answer
to this question is not as straightforward as we might naively
think. In Einstein gravity, a damping term appears explicitly into the
equations of motion for a test particle just after some type of
approximation is performed. The exact form of the dissipation term
depends not only on the coordinate system chosen, but depends also on
the approximation technique used. This is a consequence of the
non-linearity of the equations of motion. 
Moreover, the problem of gravitational
radiation reaction usually involves enormously complicated
calculations and are full of potential sources of errors which may
lead to results whose physical meaning is difficult to be established.

On the other hand, the analogous problem of the electromagnetic
radiation reaction is far easier to be analyzed and quite well
understood. 
Much work on the subject has been done since the pioneering papers
by Lorentz \cite{lorentz} and Planck \cite{planck}. The relativistic
version of the radiation reaction force was derived by Abraham
\cite{abraham} and lately by Dirac \cite{dirac}, and we
can say that the effects of radiation reaction force on an accelerated
particle, as a classical field theory in special relativity, is very well
understood (see, e.g., \cite{jackson} and references therein).
The generalization of Dirac's result to curved spacetimes was done by
DeWitt and Brehme \cite{dewittbrehme}, and by Hobbs \cite{hobbs}.
When considering the quantum theory, the classical electromagnetic radiation
reaction force is also soundly based, since it can be obtained  by
taking the appropriate limit of a particular quantum
electrodynamical process \cite{moniz}. For instance, the position of a
linearly accelerated charged particle in the Lorentz-Dirac theory is
reproduced by the $\hbar \longrightarrow 0$ limit of the one-photon emission
process in QED (See \cite{higuchi} and Refs. therein).
However, the study of chaotic radiating electromagnetic systems found in
the literature refers mostly to quantum properties of such systems.
Its classical counterpart was not investigated perhaps because the radiation
reaction is really important in microscopic systems.

The similarity between the Abraham-Lorentz theory and the equations 
appearing in some approximation schemes from the general relativistic
analogous problem of a radiating gravitational system (see e.
g. Ref. \cite{pfenningpoisson}),
and the simplicity of the electromagnetic case compared to the
gravitational case, makes interesting to deepen the study on this subject.
Therefore, we perform here the analysis of the effects of
radiation reaction forces considering a classical electromagnetic
chaotic system,
and in a future work we investigate the gravitational case. We expect that
the comparison of the results from the present work to future works
considering gravitational systems, although different in
characteristics, shed some light helping to
better understand the gravitational radiation damping problem,
particularly in chaotic systems (see e.g. \cite{kunze} for the
comparison among electromagnetic and gravitational non-chaotic damped
systems).

\subsection{The problem}

In order to investigate the effects of radiation reaction on 
the dynamics of an electromagnetic chaotic system,
we consider a charged test particle (it can be a macroscopic
test particle) of mass $m$ and charge $q$ submitted to an external
 electrostatic field. In such a case, the non-relativistic
 motion of the test particle is governed by the equation 
\cite{ford1,rohrlich}
\begin{equation}
m\frac{d\vec{v}}{dt}=\vec{F}_{\rm ext}+\tau_{\rm q}
\frac{d\vec{F}_{\rm ext}}{dt},          \label{eq:damping}
\end{equation}
where $\vec{F}_{\rm ext}$ denotes the external force acting on the
charged particle, $\displaystyle{d\vec{F}_{\rm ext}\over dt}$ is the
complete (convective) time derivative of the external force, and
$\tau_{\rm q}=2q^2/(3mc^3)$ is the characteristic dissipation time,
which indicates how efficient is the energy emission. The last term of
the above equation is the particle self-force which arises due to the
emission of electromagnetic radiation, and is interpreted as a
dissipative force. Accordingly,  such a term is usually referred to 
as  a {\it damping term},
and also as a {\it dissipation term}, both of which  are used
throughout this paper.

Accordingly, the names damping term, or dissipation
term are also used.

The derivation of Eq. (\ref{eq:damping}), and of its relativistic
version, with some applications and with the interpretation of the
dissipation term (and, in particular, of the parameter $\tau$) 
can be found in the classical textbooks
\cite{landaulifshitz,jackson}.  In fact, in the original derivation by
Lorentz \cite{lorentz} and Planck \cite{planck} (the relativistic
version was derived by Abraham \cite{abraham} and Dirac \cite{dirac})
the resulting equation of motion is $
m \, \vec a  =\vec{F}_{\rm ext}+m\tau_{\rm q}\,
(d\vec{a} /dt) $, which leads to runaway solutions. A way
to avoid such a type of solutions is by replacing 
the time derivative of the particle's acceleration $d\vec a/dt$ by 
the time derivative of the external force,
$m (d\vec a /dt)= {d\vec F_{\rm ext}/ dt}$, into this
equation, what yields  Eq. (\ref{eq:damping}) as a first approximation
to the equation of motion for a charged particle. A deeper
analysis, however, performed in Ref.  \cite{rohrlich} claims that Eq.
(\ref{eq:damping}) is the correct equation of motion for a charged
particle submitted to an external force $\vec F_{\rm ext}$ (see also
Refs. \cite{ford2,vogt}).

 A further well known property of Eq. (\ref{eq:damping}) is that, for
motions within a time interval $\Delta t$ such that $\Delta t\ll
\tau_{\rm q}$, the radiative effects on the dynamics of the system
will be negligible, and the last term in Eq. (\ref{eq:damping}) can be
neglected. Thus, in order for the effects of the damping term to be
noticeable, the time of observation must be large compared to
$\tau_{\rm q}$.
This is equivalent to say that the effects of dissipation
will be important only
for situations in which the external force is applied for a time
interval $\Delta t$ much larger than the dissipation time $\tau_q$,
$\Delta t\gg\tau_q$. These conditions were both taken into account
in our simulations (see Sec. \ref{units}).
Hence, the system we
are analyzing here can be interpreted as the analogous to the case of an
orbiting test body in a weak gravitational field, but considering
explicitly the damping term.

\subsection{The structure of the paper}

In the following section we write explicitly the equations of motion
for a test charged particle in the H\'enon-Heiles potential, by
assuming a non relativistic motion.  Sec. \ref{sectnumerical} is
dedicated to report the numerical results and to their analysis. A
brief analysis on the relativistic particle motion is done in
Sec. \ref{sectrelativistic}, and finally in Sec. \ref{sectfinal} we
conclude by making a few remarks and final comments.

\section{H\'enon-Heiles Systems}
\label{sectthemodel}

\subsection{The model}

We consider an external force $\vec F_{\rm ext}$ derived from a
H\'enon-Heiles electrostatic potential \cite{hh}, and work in a
non-relativistic regime where Eq. (\ref{eq:damping}) holds (for a
relativistic version of Eq. (\ref{eq:damping}) see
\cite{rohrlich,ford2}; see also Sec. \ref{sectrelativistic}).  The
choice of such a potential was due mainly to its simplicity allied to
its dynamical richness, implying for instance chaotic motions, what is
of capital importance in our analysis. Other interesting point to be
mentioned is that a potential of the same type was used in a previous
work which analyzed the emission of gravitational waves \cite{kokubun}
instead electromagnetic waves, and so the results of the two works can
be compared.  H\'enon-Heiles systems are described by a potential of
the form
\begin{equation}
V(x,y)=m\omega^2\frac{x^2+ y^2}{2}+ \frac{m\omega^2}{a}\left(x^2 y
-\frac{y^3}{3}\right),
\label{eq:potHH}
\end{equation}
and have been considered in several contexts \cite{vernov} beyond the
original astrophysical scenario. This potential is basically a
perturbed two-dimensional harmonic oscillator. Therefore, $\omega$ may
be identified with the oscillatory frequency which, in the absence of
the perturbation term, is $\omega=\sqrt{k/m}$, $k$ being a spring
constant, for a mechanical system or $\omega=\sqrt{|Qq|/ma^3}$, where
$Q$ and $q$ are respectively the source and the test particle charges,
for an electric system (in CGS-Gaussian units). Parameter $a$ is the
characteristic length of the system. The characteristic frequency
$\omega$ defines a characteristic period of motion, $T=\omega^{-1}$.

Without the damping term, and with the usual choice of units
\cite{hh}, $m=\omega=a=1$, and in our case also $Q= q=1$ (see below),
the chaoticity of the  H\'enon-Heiles system is controlled only by its
energy $E$: the system is bound if $E\lesssim 1/6$, being mostly
regular for the energy range from $0$ to nearly $1/10$, and
being mostly chaotic for $E$ in the range $1/10$ to $1/6$.

In the presence of the damping term, the dynamics of a charged point
particle in the potential given by equation
(\ref{eq:potHH}) is governed by the equations
\begin{eqnarray}
\ddot x+\omega^2 x &=&-2\frac{\omega^2}{a}xy-\tau_q\omega^2\left[\dot{x}+
2\frac{1} {a} \left( x\dot y + y\dot x\right) \right]\,
,\label{xmotioneq}\\ 
\ddot y + \omega^2 y&=& -\frac{\omega^2}{a}
\left(x^2-y^2\right) -\tau_q\omega^2\left[\dot{y}+2\frac{1}{a}
\left(x\dot x - y\dot y\right)\right]\, . \label{ymotioneq}
\end{eqnarray}

Here, working with electromagnetic field and using
Eqs. (\ref{xmotioneq}) and (\ref{ymotioneq}), we considered the
effects of radiation damping, comparing long term energy loss between
chaotic and regular regimes. The energy loss being considered directly
into the equations of motion without further approximations.  The main
results are reported and analyzed in 
Sec. \ref{sectnumerical}.

\subsection{Units and normalized parameters}
\label{units}

We present here a discussion about the physical parameters of the
present H\'enon-Heiles electromagnetic system. However, let us
stress once more that the present model is to be considered a toy 
model, as a laboratory test for our procedures, and not as 
a test for the electromagnetic theory.

Eqs. (\ref{xmotioneq}) and (\ref{ymotioneq}) have three free
parameters characteristic to the system under consideration: The
constant $a$, the characteristic time $\tau _{\rm q}$, and the
frequency $\omega$.  Then we follow the standard procedure and choose
a new normalized time parameter $\tau$ and a new normalized time
variable $t$ given respectively by the relations $\tau = \tau_{\rm
q}\times \omega$, and $t= t(s)\times \omega$, where $\tau_{\rm q}$ and
$t(s)$ carry dimensions, while $\tau$ and $t$ are dimensionless
parameters.  The constant $a$, which carries dimensions of length, is
used to normalize the variables $x$ and $y$. The usual choice is to
measure $x$ and $y$ in units of $a$, which is equivalent to making
$a=1$ into the equations of motion.

As far as the effects of dissipation are concerned, the important
parameter is the rationalized characteristic time $\tau$. The
contribution of the radiation reaction force to the dynamics of the
system is proportional to $\tau$ (see Eq. \ref{eq:damping}). Therefore, 
the value to be chosen
for $\tau$ has to be as large as possible. On the other hand, as shown
below, the time of observation (the computation time) has to be much
larger than $\tau$ in order for the effects of dissipation being
noticeable.

Considering the motion of charged elementary particles, the largest
value for $\tau _{\rm q}$ follows when the test particle is an
electron, in which case one has $\tau_{\rm q} \equiv\tau_{\rm e}\simeq
6.3\times 10^{-24}s$. If the test particle is a proton then
$\tau_{q}\simeq 3.4\times 10^{-27}s$. For macroscopic systems,
however, the ratio $q/m$ is not fixed and $\tau_{q}$ may assume values
several orders of magnitude larger than $\tau_{\rm e}$. For a charged
test body such that $q=\alpha e$ and $m=\beta m_{\rm e}$ one has
$\tau_{\rm q}= (\alpha^2/\beta) \tau_{\rm e}$. Take, for instance,
a test body of mass $m= 10 g$ and charge 
$q=1.0\times 10^{7}\, \rm{esu} \simeq
3.3\times 10 ^{-3}C$. Then, it follows $\alpha = 2.1\times 10^{16}$,
$\beta =1.1 \times 10^{28}$, and $\tau_{\rm q}= 2.5\times 10^{-19}s$.

Regarding to the third parameter of the model, the characteristic
frequency $\omega$, one sees that it depends also upon the source of
the H\'enon-Heiles potential, being a typical period of the system.
For an electromagnetic system it is related to the total charge $Q$ of
the source by a relation of the form, $m\omega^2 = Qq/a^3$, 
$m$ and $q$ being respectively the mass and the
electric charge of the test particle, and $a$ being the characteristic
length of the system mentioned above. As usual in H\'enon-Heiles
systems, we are free to fix $\omega$ as an inverse time unit,
$\omega=\omega_q$, in such a way that $\tau = \tau_{q}\omega_q$.
Thus, if the source of the potential has mass $m=\beta m_e$
and net electric charge $q=\alpha e$, we have
\begin{equation}
\tau=\tau_q\omega_q=\frac{\alpha^2}{\beta}\tau_e\omega_q\, .
\label{dissipationtime}
\end{equation}
Therefore, if we think of a specific orbiting particle ($\tau_q$ fixed), 
different values of the dissipation parameter mean different values
for $\omega_q$, which gives the corresponding physical parameters of the
H\'enon-Heiles potential. 

Now we are ready to fix the parameters and to establish some constraints
to the physical size of our system.  Let us consider a system with
typical size $L$ (which is essentially of the order of the
parameter $a$ mentioned above). Being $\omega_q^{-1}$
a typical period, we have that $L\omega_q$ is a typical velocity of
the system. Such a velocity has to be smaller than the speed of light
$c$, i.e., $L\omega_q\le c $, so that we have the following upper bound
for the system size

\begin{equation}
L \lesssim c \omega_q^{-1}= c
\frac{\tau_e}{\tau}\frac{\alpha^2}{\beta}\simeq
2\times10^{-13}\frac{\alpha^2}{\tau\beta}\;\rm{cm}. \label{systemsize}
\end{equation}

As we shall see below, physically interesting values of $\tau$ for
which the numerical results can be clearly interpreted lay in the
interval $\tau \in [  10^{-10},  10^{-4}]$.  Considering
such a range for $\tau$ and taking an electron as the test particle,
for which $\alpha=\beta=1$, we obtain the typical size of the system
as $L\lesssim 2\times 10^{-9}\rm{cm}$ for $\tau =   10^{-4}$,
and $L\lesssim 2\times 10^{-3}\rm{cm}$ for $\tau =   10^{-10}$.
On the other hand, if the test body has mass $m=10g$ and charge
$q=  10^7\rm{esu}$ ($\alpha = 2.1\times 10^{16}$, $\beta =1.1
\times 10^{28}$), we obtain respectively, $L\lesssim 8\times 10^{-5}
\,\rm{cm}$ and $L\lesssim 80\rm{cm}$. 
If the test particle is an
electron, the typical size of the system is microscopic which is more
difficult to be managed. Thus, in order to consider a possible
experimental setup, it will be certainly more feasible to work with a
macroscopic test particle.  However, the choice of $\tau$ in the above
range describes both the microscopic and the macroscopic systems.

 One more issue on the subject of fixing parameters concerns
the physical properties of the source in the H\'enon-Heiles system.
Namely, the electric charge $Q$ and the characteristic length $a$.
As we have seen above, these are
related to the characteristic frequency $\omega_q$, and once we have
normalized units through Eqs. (\ref{dissipationtime}) and
(\ref{systemsize}), the ratio $Q/a^3$ is fixed as soon as we fix the
dissipation parameter $\tau$. From the above definitions and choices
it is found $Q \sim 1.7\times 10^{3}\beta\,
L/\alpha\,\,(\rm{CGS})$,
where we assumed that the characteristic size of the source 
is of the same order of magnitude as the parameter $L$ defined in Eq. 
(\ref{systemsize}). Hence, in the case of the preceding examples
it gives the upper limits $Q\sim 3.6 \,\rm{esu}$ and $
3.6\times 10^{-6}\,\rm{esu}$ for the microscopic orbiting particle,
respectively with $\tau=10^{-10}$ and $\tau=10^{-4}$. And for the
macroscopic orbiting particle we find 
$Q\sim 7.1\times 10^{16}\,\rm{esu}$ and $ 7.1\times 10^{10}\,\rm{esu}$,
respectively, for $\tau=10^{-10}$ and $\tau=10^{-4}$.

\section{Numerical simulations and results}
\label{sectnumerical}

\subsection{Methodology}

Equations (\ref{xmotioneq}) and (\ref{ymotioneq}) were solved
numerically. For the sake of comparison, we initially used two
different numerical methods: a fourth order Runge-Kutta with fixed
stepsize and a Runge-Kutta with adaptive stepsize \cite{NR}. Also, we
used MATHEMATICA built-in procedures for solving Ordinary Differential
Equations.

At first, we integrated Eqs. (\ref{xmotioneq}) and (\ref{ymotioneq})
considering no dissipation term, i.e., with $\tau=0$. The initial
conditions were generated at random, fixing only initial energy and
choosing $x=0$ at the start. Since in this case the system is conservative,
the energy
is a constant of motion, and its value was used to check and compare
the numerical results obtained through different methods.  No
significant differences were observed, so we adopted a fourth order
Runge-Kutta in our simulations.

\subsection{Numerical simulations}

The main concern of this work is answering the question: How does the
value of the dissipation time $\tau$ affect the dynamics of system?
In particular, we also want to know how much energy is radiated by the 
accelerated particle undergoing chaotic motions in comparison to regular
motions.
It is expected that
a large value of $\tau$ will strongly affect the dynamics, in
opposition to small values, for which the dynamics of the system
should be weakly affected. Nonetheless, it remains to be defined what
values of $\tau$ can be considered large and what are small ones.  For
comparison we made simulations for several different values of $\tau$
and same initial conditions. After a few tries, we have chosen five
particular cases to analyze in more detail.  The chosen values are
$\tau=0,\;10^{-4},\;10^{-6},\;10^{-8}$, and $10^{-10}$, with initial
energy $E=0.12$ and the same initial conditions for all of the five cases.

\begin{figure}
\includegraphics[scale=0.3]{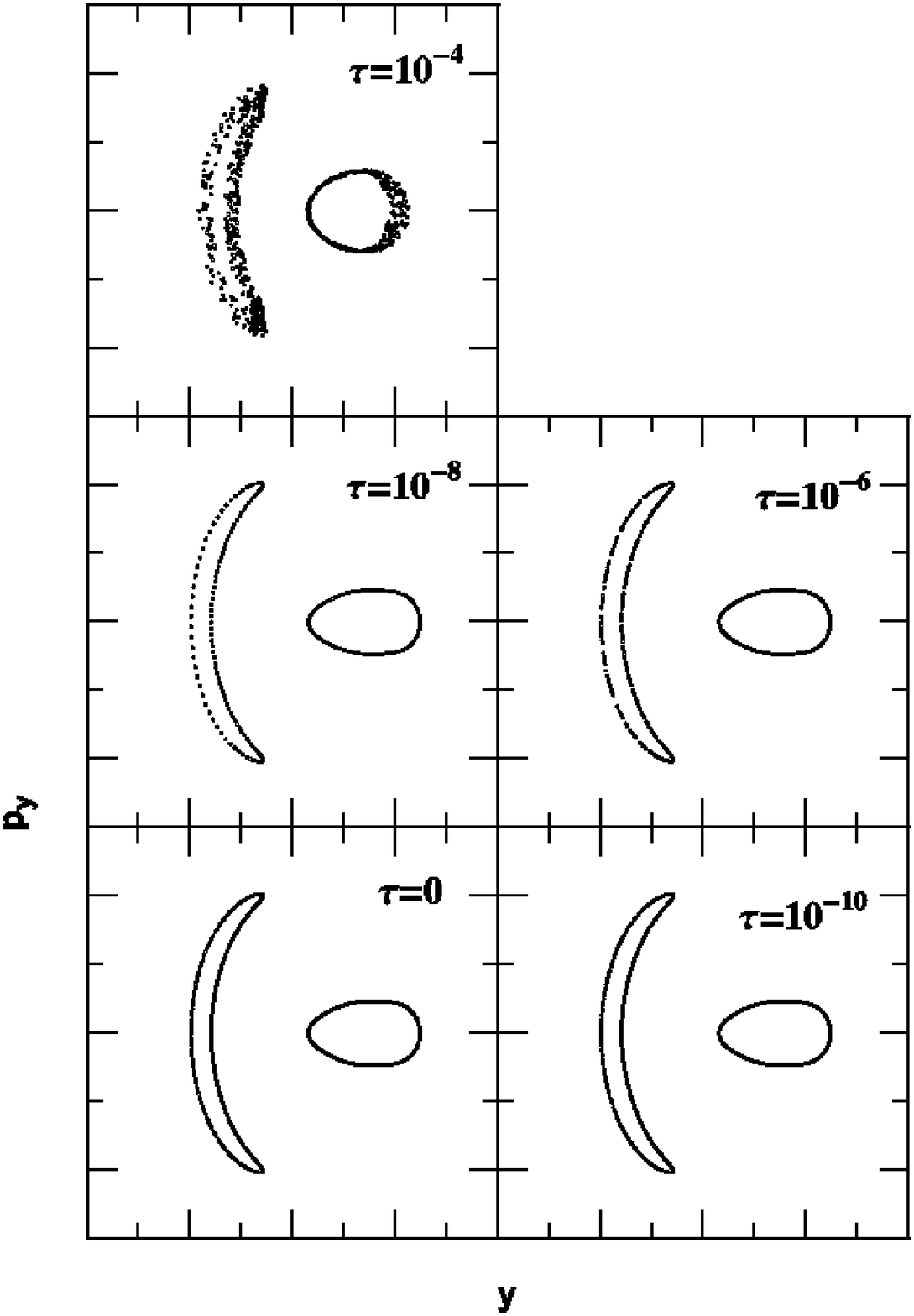}
\caption{Poincar\'e sections for $E=0.12$ and several values of
$\tau$. All sections with the same initial
conditions.}\label{fig:sos12b_taus}
\end{figure}

The results can be seen in Fig. \ref{fig:sos12b_taus}, where we plot
the Poincar\'e sections for each value of $\tau$. Each one of the
graphs represents the resulting section for only one orbit,
corresponding to the particular initial conditions we have chosen.
 Notice that
except for $\tau=10^{-4}$, all other Poincar\'e sections look very
similar, suggesting that $\tau \sim 10^{-4}$ or greater are to be
considered large values, and $\tau$ is to considered small if its
value is of the order of $10^{-6}$ or below.

\begin{figure}
\includegraphics[scale=0.4]{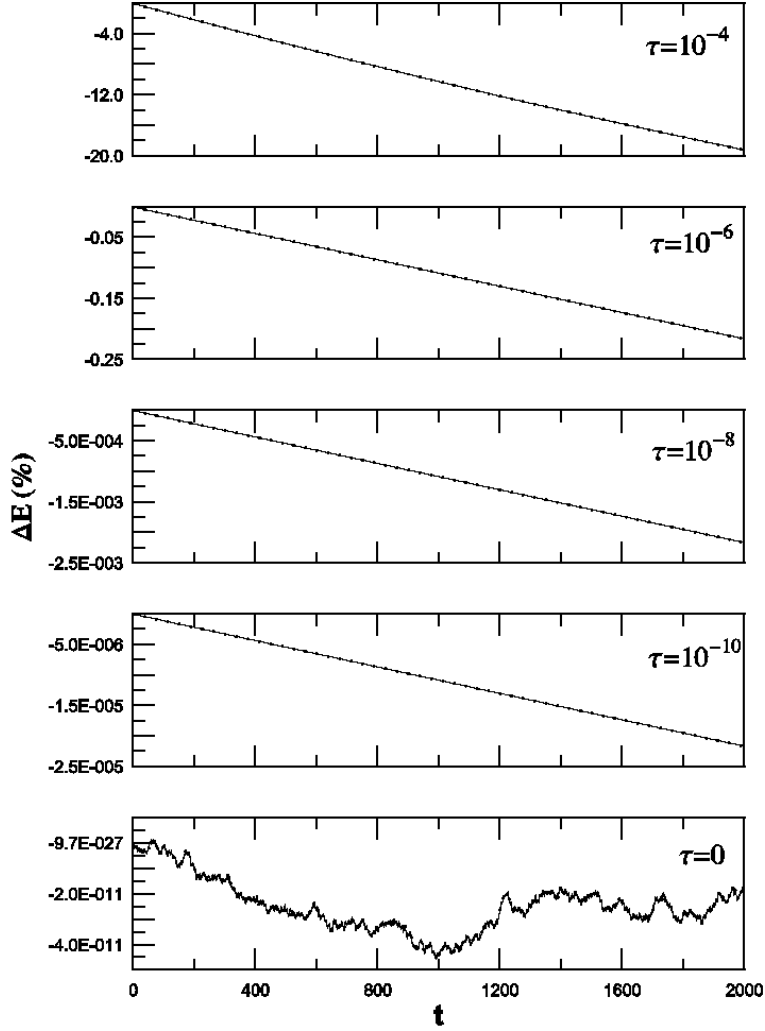}
\caption{Percentage of energy loss for five values of
$\tau$, as indicated in the figure.
Initial energy is E=0.12, and the same initial conditions
were used for all cases, as in Fig. \ref{fig:sos12b_taus}.}
\label{fig:de_12}
\end{figure}

\begin{figure}
\includegraphics[scale=0.4]{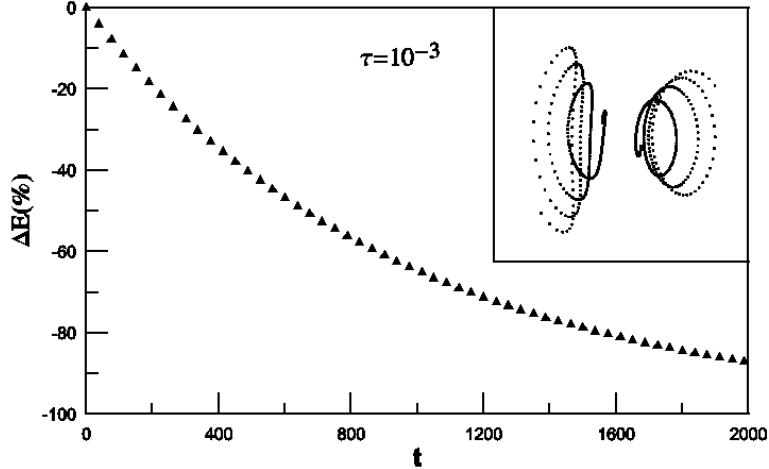}
\caption{Energy emission with high $\tau$ term as a function of time.
The small graphics is the Poincar\'e section of the motion. The initial
energy is $E=0.12$ and $\tau=10^{-3}$.}\label{fig:hightau}
\end{figure}

In order to better understand the behavior of the Poincar\'e sections we
evaluate the percent amount of radiated energy $\Delta E$ as a function
of time in each case shown in Fig. \ref{fig:sos12b_taus}
The results are seen in Fig. \ref{fig:de_12} where we plot the graphics of
$\Delta E\times t$ for each case. 
Note that these graphs are the actual data points,
and not fits adjusting the data. For instance, the lines 
appearing in the first four graphs of that figure are the result of
plotting the set of points obtained numerically for each one of
the particular orbits chosen to be analyzed. Such straight lines
indicate that energy emission rate is constant, and that
the total energy of the system decreases linearly with time.
This is so for small dissipation times $\tau$, while for higher values
of $\tau$ the energy loss rate is not constant with time 
(see Fig. \ref{fig:hightau}).
It is seen that for $\tau=0$ the variations in
the energy are exceedingly small ($\sim 10^{-11}\, \%$) and look like
random variations. This is surely not an effect of dissipation,
because the total energy dissipated during the integration time is
essentially zero.  These random variations are caused by numerical
inaccuracy, as it can be inferred by comparing this to the other cases
with $\tau\neq 0$, where the energy variations are much larger and
systematic, causing the energy to decrease monotonically with time.
For instance, for $\tau=10^{-4}$ the total energy dissipated during
the integration time reaches nearly $20\,\%$ of the initial value, so
that at time $t \simeq 2000$ the energy of the system is about
$E\simeq 0.096$. On the other hand, for $\tau=10^{-6}$ the energy
variation reaches nearly $0.20\%$ of its initial value in the same
integration time, and the energy is reduced to nearly $E\simeq
0.1197$, meaning it is almost a constant of motion. Also for
$\tau=10^{-8}$ and $10^{-10}$ the energy variations reach $\sim
10^{-3}\,\%$ and $ \sim 10^{-5}\,\%$, respectively.  Even though these
energy variations are quite small, they are about eight (for $\tau=10^{-8}$)
and six (for $\tau=10^{-10}$) orders of magnitude larger then in the
case $\tau=0$, and yet we can see they cause the energy of the system
to decrease monotonically with time.
  
As a further example of a large (higher) $\tau$ value, we performed
simulations with $\tau=10^{-3}$ and with energy $E=0.12$, and the results
are seen in Fig. \ref{fig:hightau}.  The large graphics shows the percent
variations of the energy, and the small graph is the Poincar\'e
section for this special orbit.  Due to the large dissipation parameter,
the motion is highly damped and, after some time, rest is attained. At time
$t=2000$ nearly $90\%$ of the initial energy has been carried away by
electromagnetic radiation. Note also that the amount of energy emitted 
does not vary linearly with time, as it happens for smaller values of $\tau$.

For our purposes here, high values of $\tau$, $\tau  \gtrsim 
10^{-4}$, say, are not interesting because  the dynamics
of the system is highly affected and the comparison to the case
without dissipation becomes difficult (if not impossible) to be done.
Then, we considered only $\tau= 10^{-6},\;\tau=10^{-8}$, and $\tau=10^{-10}$
in our full simulations.

Once fixed the values of $\tau$, the next step was solving numerically
Eqs. (\ref{xmotioneq}) and (\ref{ymotioneq}) with several 
initial conditions, and for the energy values $E= 0.12$ and
$E=0.14$. A set of $500$ distinct initial conditions was 
generated using a (pseudo) random number generator \cite{RN}, and the
same set was used for every combination of the 
controlled parameters, $\tau \in \{0,\; 10^{-6},\; 10^{-8},\;
10^{-10}\}$  and $ E \in \{0.12, \;0.14\}$. 
For each pair of these parameters we performed $500$ simulation,
with the $\tau=0$ case being included only for comparison purposes.

Although we performed simulations also for $E=0.09$ and $E=0.10$, the
respective results were not considered in our analysis.  In such
cases, the number of chaotic motions (typically less than $10$ in
$500$ simulations) in our set of results was too small for a good
statistics, and so they would not be useful in comparing chaotic to
regular regimes, which is the basic aim of the present work.

\subsection{Results and analysis}
\label{resultssect}

\begin{figure}
\includegraphics[scale=0.5]{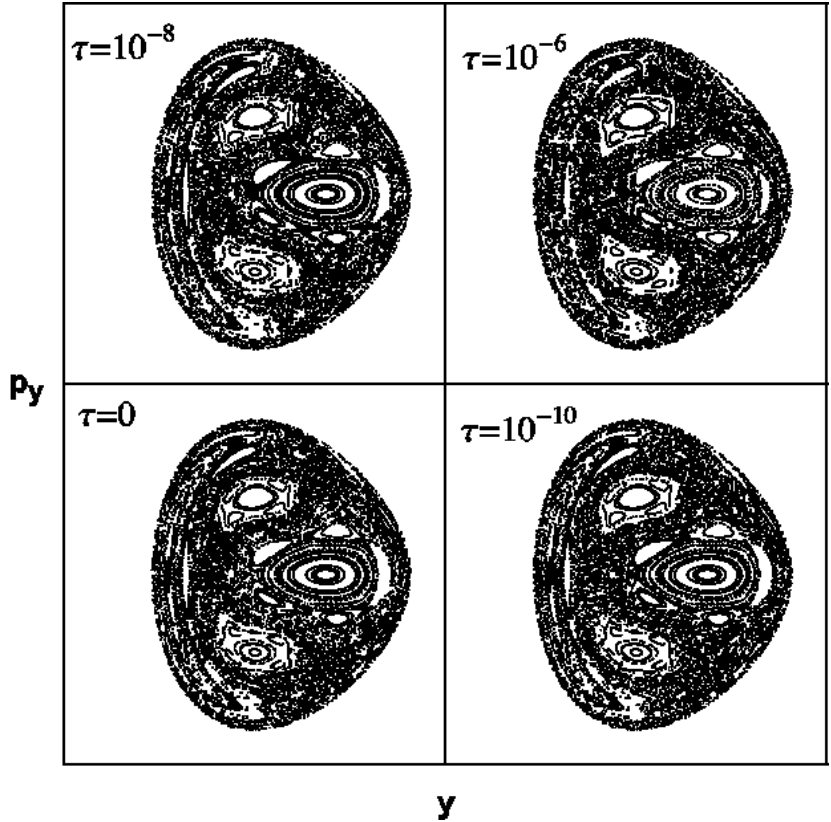}
\caption{Poincar\'e sections for $E=0.12$ and 
four  values of $\tau$, as indicated in each graphics,
corresponding to $500$ distinct initial conditions in each plot.}
\label{fig:poincare}
\end{figure}

Using the results of our simulations, we constructed Poincar\'e
section for each case, all of them being drawn on the surface
$x=0$ in phase space. These sections were analyzed in order to
separate between dynamics with chaotic motions from 
dynamics with regular motions.

\begin{figure}
\includegraphics[scale=0.4]{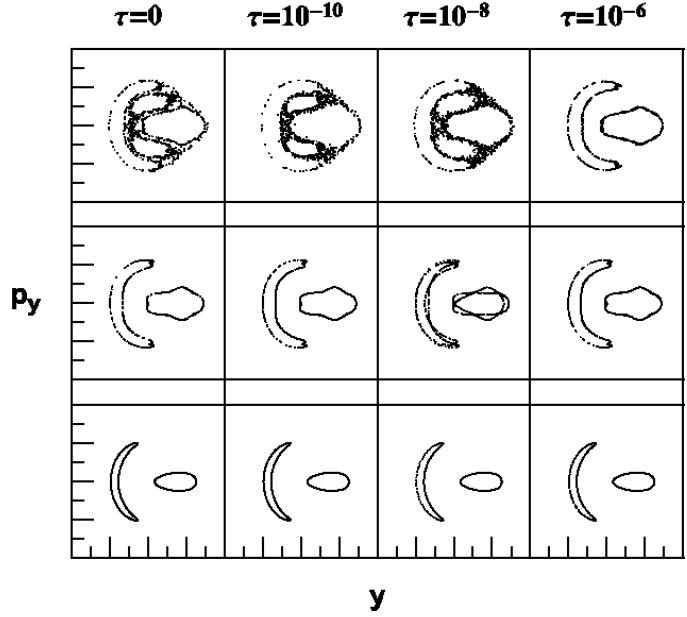}
\caption{Poincar\'e sections for $E=0.12$ and four distinct values of
$\tau$, but with the same initial conditions, are shown at the same
horizontal level. The corresponding 
$\tau$ values appear on top of each column.}
\label{fig:sos_comparando}
\end{figure}

The graphics in Fig. \ref{fig:poincare} are Poincar\'e sections
obtained for $E=0.12$, without damping term ($\tau=0$), and with
dissipation term for $\tau= 10^{-10}$, $\tau=
10^{-8}$, and $\tau=10^{-6}$, as indicated in each plot.
 Although the overall aspects are the same, a detailed
analysis of individual sections reveals different aspects as seen in
Fig. \ref{fig:sos_comparando}, where we plot Poincar\'e sections for
particular orbits corresponding to three different sets of initial
conditions for each value of $\tau$.   Here a very important result is
that with the inclusion of a small damping term, the overall aspect of
motions are the same, and in particular, the energy is nearly
constant, so that Poincar\'e section is still a good tool in order to
classify the motions as regular or chaotic. From this figure it is
also seen the dependence of the dynamics on the initial conditions,
besides the dependence upon $\tau$.  It is also worth saying that a few
particular orbits, out of the 500 initially chosen, were neglected since
it was not clear from the obtained Poincar\'e sections whether they
correspond to regular or chaotic motions (see Table \ref{tab:deltaE}).

Now with our set of simulations already separated into two sets, one
set with ordered motions only and the other set with chaotic ones, we
calculated for each initial condition (in each set) a best linear fit to
the energy variations $\Delta E = A t + B$, determining $A$ and $B$ by
using standard techniques of linear regression. Then, with such a set
of values for $A$ and $B$ we determined the mean values $<A>$ and $<B>$,
and their respective standard deviations, $\sigma_A$ and $\sigma_B$,
for each regime of motions.  The results are summarized in Table
\ref{tab:deltaE}, where we show just the values of
$<A>$ and $\sigma_A$. These are more important than the values of $<B>$
and $\sigma_B$, because they furnish the (time) rate of energy carried
away by electromagnetic waves. We also show in that table
(last column), the resulting number of regular and chaotic motions for
each pair of values of the initial energy and dissipation time. As
mentioned above, some orbits are missing because they could not be
classified as regular nor as chaotic ones.


\begin{table}
\begin{tabular}{|c|r|}
\hline 
$E=0.12 $ & {\begin{tabular}{|c|p{2.25cm}|p{2.08cm}|p{1.32cm}|p{.58cm}|} 
 $\tau$ & $ \qquad<A>$ & $\qquad\sigma_A$ &  Regime & $\;\# $\\
\hline
  $10^{-6}$ & $-9.92\times 10^{-5}$  $-9.24\times 10^{-5}$    
   & $1.33\times 10^{-5} \;$  $3.99\times 10^{-6} $
&Regular Chaotic  & 282  202 \\\cline{1-5}
 $10^{-8}$ & $-1.02\times 10^{-6}$ $-9.18\times 10^{-7}$ 
 & $1.29\times 10^{-7} \;$ $5.39\times 10^{-8}$ 
&Regular  Chaotic & 236  260\\ \hline
  $10^{-10}$ & $-1.08\times 10^{-8}$ $-9.09\times 10^{-9}$
& $9.09\times10^{-10}$ $6.90\times10^{-10}$ 
 &Regular Chaotic  & 163 337\\  \end{tabular}}\\
\hline
  $E=0.14$ &  {\begin{tabular}{|c|p{2.25cm}|p{2.08cm}|p{1.32cm}|p{.58cm}|}
  $10^{-6}$ & $-1.05\times 10^{-4}$ $-9.06\times 10^{-5}$
& $1.53\times 10^{-5} \;$ $4.76\times 10^{-6} $ 
&Regular Chaotic  & 146  350 \\ \hline
 $10^{-8}$ & $-1.06\times 10^{-6}$ $-9.05\times 10^{-7}$ 
& $1.51\times 10^{-7}\;$ $5.28\times 10^{-8} $
&Regular Chaotic & 141  359 \\ \hline
 $10^{-10}$ & $-1.12\times 10^{-8}$ $-9.06\times 10^{-9}$
& $7.97\times 10^{-10}$ $7.42\times 10^{-10} $ 
&Regular  Chaotic& $\;96$ 403\\ \end{tabular}}\\
\hline
\end{tabular}
\caption{Mean values of the  energy loss rates $<A>$, and the standard deviations
$\sigma_A$ for each energy and $\tau$ values. The last column shows
the number of motions in each particular regime.}
\label{tab:deltaE}
\end{table}

With the obtained data, we compared the amount of energy radiated in
regular regimes with respect to chaotic regimes, and calculated the
percent ratio $\eta$ as follows
\begin{equation}
\eta= \frac{<A>_R - <A>_C}{<A>_C} \times 100, 
\end{equation}
where the subscript $C$ stands for chaotic and subscript $R$, for regular.
This is shown in Table \ref{tab:ratio}. In all
cases the average energy radiated in regular regimes is 
larger than in chaotic regimes. These results are compatible with what was
obtained when considering gravitational waves emission
\cite{kokubun,suzuki} (see also \cite{levin2000,cornish}).

\begin{table}
\begin{tabular}{|c|c|c|} \hline
$E$ & $\tau$ & $\eta$ (\%) \\ \hline 
$0.12$ & $10^{-6}$ &7\% \\ 
0.12 & $10^{-8}$ &11\% \\
0.12 & $10^{-10}$& 18\% \\ \hline 
0.14 & $10^{-6}$ &16\% \\ 
0.14 & $10^{-8}$ &17\% \\
0.14 & $10^{-10}$ &24\% \\ \hline
\end{tabular}
\caption{The percent values of energy emitted by the system in
a regular regime with respect to a chaotic regime.}
\label{tab:ratio}
\end{table}

\section{Relativistic motion}
\label{sectrelativistic}

We have also investigated the behavior of the electromagnetic
H\'enon-Heiles system in relativistic dynamics. In such a case we
solved the equations \cite{rohrlich, jackson}
\begin{eqnarray}
{d\vec p \over d t} = \vec F + \tau\gamma {d\vec F\over dt} -
\tau{\gamma^3\over c^2} {d\vec v\over dt}\times\left(\vec v \times\vec
F \right) \, ,
\label{eq:relat}
\end{eqnarray}
where $\vec p = \gamma m \vec v$, $\gamma = 1/\sqrt{1-v^2\,}$, and
$\vec F$ is the external force given by $\vec F =-\vec\nabla U$, U
being the potential function given by
Eq. (\ref{eq:potHH}).  The explicit form of Eqs. (\ref{eq:relat}), 
analogous to Eqs. (\ref{xmotioneq}) and (\ref{ymotioneq}), 
were used in the numerical calculations.

The numerical results obtained from the relativistic equation
(\ref{eq:relat}) were essentially the same as in the non-relativistic
case.  This can be understood noticing that for the bound system the
particle undergoes a non-relativistic motion, as can be verified by
the following facts. In the H\'enon-Heiles potential (\ref{eq:potHH}),
for the test particle to acquire velocities comparable to the velocity
of light, its initial energy $E$ has to be large. In the rationalized
units used here, this means $E\sim 1$. However, as shown in
Ref. \cite{hh}, if $E$ is larger than $1/6$ the system is not bound,
and then in the regime where relativistic effects become important the
particle is not bound by the H\'enon-Heiles potential.  Therefore, the
relativistic regime is not important in the present analysis.

\section{Final remarks}
\label{sectfinal}

Our results show that when we consider explicitly the effects of
radiation reaction force, the energy emission through electromagnetic
waves in the chaotic regime is smaller than in the regular regime, as it
was  in the case of emission of gravitational waves. 

The ratio of energy loss in regular compared to
chaotic motions increases with the initial energy of the system,
and decreases with the dissipation parameter. Since in H\'enon-Heiles
systems the chaoticity increases with the energy, this means that 
the ratio between the energy emitted in regular motions and in chaotic
motions grows with the chaoticity of the system.

We recall
that, in the gravitational waves case, the simulations were performed
at PN approximation lower than 2.5PN. The result was that the effect
of gravitational waves emission is negligible to the dynamics of the
system. However, being PN lower than $2.5$, in those simulations the
effects of radiation emission were in fact not fully considered. In
our analysis of the electromagnetic H\'enon-Heiles system, these
effects are fully considered through the radiation reaction force.
Another
important result is related to the mean life-time of source. If we
make a prediction considering only regular dynamics its mean life-time
may be shorter than the prediction from chaotic dynamics.  
However, in the case of
dissipation by emission of gravitational radiation a more careful
analysis has to be done.

The numerical procedures and analysis performed in this work will
certainly be useful in our task of studying the gravitational
analogous problem, the one about the gravitational radiation emitted by
a particle undergoing chaotic motion, considering explicitly
the damping term into the equations of motion 
(work on this subject is in progress).

\section*{Acknowledgments}
 
This work was partially supported by Funda\c{c}\~ao de Amparo \`a
Pesquisa do Estado do Rio Grande do Sul (FAPERGS). 
We thank A. S. Miranda and S. D. Prado for useful
conversations.

\end{document}